\documentclass{appolb}
\usepackage{epsfig,amssymb,floatflt,amsmath}
\newcommand{\as}{\alpha_\mathrm{s}}
\newcommand{\pom}{\mathrm{I\!P}}
\newcommand{\cV}{\mathcal{V}}
\newcommand{\cP}{\mathcal{P}}

\newcommand{\NP}{\mathrm{NP}}
\newcommand{\PT}{\mathrm{PT}}
\newcommand{\ee}{e^{+}e^{-}}
\newcommand{\order}[1]{{\mathcal{O}\left(#1\right)}}
\begin{document}

\title{
\mbox{ }\vspace{-4cm}
\begin{flushright}
  \textrm{
  LPTHE--P02--04\\
  hep-ph/0207147 \\
  July 2002\vspace{1.4cm}}
\end{flushright}
QCD tests through hadronic final-state measurements%
\thanks{Plenary presentation at the X International Workshop on Deep
    Inelastic Scattering (DIS2002) Cracow, Poland.}%
}
\author{
G.~P. Salam
\address{LPTHE, Universit\'es Paris VI et 
  Paris VII, Paris, France.}
}
\maketitle
\begin{abstract}
  Modern-day `testing' of (perturbative) QCD is as much about pushing
  the boundaries of its applicability as about the verification that
  QCD is the correct theory of hadronic physics.  This talk gives a
  brief discussion of a small selection of topics: factorisation and
  jets in diffraction, power corrections and event shapes, the
  apparent excess of $b$-production in a variety of experiments, and
  the matching of event generators and NLO calculations.
\end{abstract}

%----------------------------------------------------------------------
\section{Introduction}

\begin{floatingfigure}[r]{0.4\textwidth}
  \mbox{ } \hspace{-0.05\textwidth}
  \epsfig{file=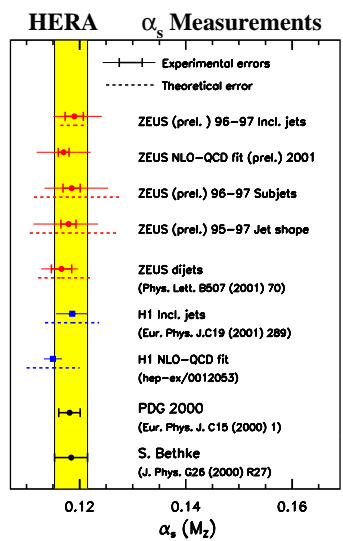,width=0.35\textwidth}
  \caption{A compilation of HERA $\as$ measurements, taken from
    \cite{ZEUSalphas}.}
  \label{fig:HERAalphas}
\end{floatingfigure}
The testing of QCD is a subject that many would consider to be well
into maturity. The simplest test is perhaps that $\as$ values measured
in different processes and at different scales should all be
consistent. It suffices to take a look at compilations by the PDG
\cite{PDG} or Bethke \cite{Bethke} to see that this condition is
satisfied for a range of observables, to within the current
theoretical and experimental precision, namely a few percent. There
exist many other potentially more discriminatory tests, examples
explicit measurements of the QCD colour factors \cite{ColourFactors}
or the running of the $b$-quark mass \cite{Bambade} --- and there too
one finds a systematic and excellent agreement with the QCD
predictions. A significant amount of the data comes from HERA
experiments, and to illustrate this, figure~\ref{fig:HERAalphas} shows
a compilation of a subset of the results on $\as$, as compiled by ZEUS
\cite{ZEUSalphas}.

In the space available however, it would be impossible to give a
critical and detailed discussion of the range of different observables
that are used to verify that QCD is `correct'. Rather let us start
from the premise that, in light of the large body of data supporting
it, QCD \emph{is} the right theory of hadronic physics, and consider
what then is meant by `testing QCD'.

One large body of activity is centred around constraining QCD. This
includes such diverse activities as measuring fundamental (for the
time being) unknowns such as the strong coupling and the quark masses;
measuring quantities such as structure functions and fragmentation
functions, which though formally predictable by the theory are beyond
the scope of the tools currently at our disposal (perturbation theory,
lattice methods); and the understanding, improvement and verification
of the accuracy of QCD predictions, through NNLO calculations,
resummations and projects such as the matching of fixed-order
calculations with event-generators. One of the major purposes of such
work is to provide a reliable `reference' for the inputs and
backgrounds in searches for new physics.

A complementary approach to testing QCD is more about exploring the
less well understood aspects of the theory, for example trying to
develop an understanding of non-perturbative phenomena such as
hadronisation and diffraction, or the separation of perturbative and
non-perturbative aspects of problems such as heavy-quark decays;
pushing the theory to new limits as is done at small-$x$ and in
studies of saturation; or even the search for and study of
qualitatively new phenomena and phases of QCD, be they within
immediate reach of experiments (the quark-gluon plasma, instantons) or
not (colour superconductors)!

Of course these two branches of activity are far from being completely
separated: it would in many cases be impossible to study the less well
understood aspects of QCD without the solid knowledge that we have of
its more `traditional' aspects --- and it is the exploration of novel
aspects of QCD that will provide the `references' of the future.

The scope of this talk is restricted to tests involving final states.
Final states tend to be highly discriminatory as well as complementary
to more inclusive measurements.  We shall consider two examples where
our understanding of QCD has seen vast progress over the past years,
taking us from a purely `exploratory' stage almost to the `reference'
stage: the question of jets and factorisation in diffraction
(section~\ref{sec:Diff}); and that of hadronisation corrections in
event shapes (section~\ref{sec:Hadr}). We will then consider two
questions that are more directly related to the `reference' stage:
the topical issue of the excess of $b$-quark production seen in a
range of experiments (section~\ref{sec:Heavy}); and then the problem
of providing Monte Carlo event generators that are correct to NLO
accuracy, which while currently only in its infancy is a subject whose
practical importance warrants an awareness of progress and pitfalls.

For reasons of lack of space, many active and interesting areas will
not be covered in this talk, among them small-$x$ physics, progress in
next-to-next-to-leading order calculations, questions related to prompt
photons, the topic of generalised parton distributions and
deeply-virtual Compton scattering, hints (or not) of instantons, a
range of measurements involving polarisation and so on. Many of these
subjects are widely discussed in other contributions to both the
plenary and parallel sessions of this conference, to which the reader
is referred for more details. 

%----------------------------------------------------------------------
\section{Jets in diffraction and factorisation}
\label{sec:Diff}

%ADD SOME SPIEL ABOUT DIFFRACTION FOR EXAMPLE IN HIGGS PRODUCTION?
%No!

Factorisation, for problems explicitly involving initial or final
state hadrons, is the statement that to leading twist, predictions for
observables can be written as a convolution of one or more
non-perturbative but universal functions (typically structure or
fragmentation functions) with some perturbatively calculable
coefficient function.

\begin{floatingfigure}[r]{0.42\textwidth}
  \mbox{ } \hspace{-0.07\textwidth}
  %\epsfig{file=schilling-feyn-diff.grey.eps,width=0.36\textwidth,clip=}\\
  %\mbox{ } \hspace{-0.05\textwidth}
  \epsfig{file=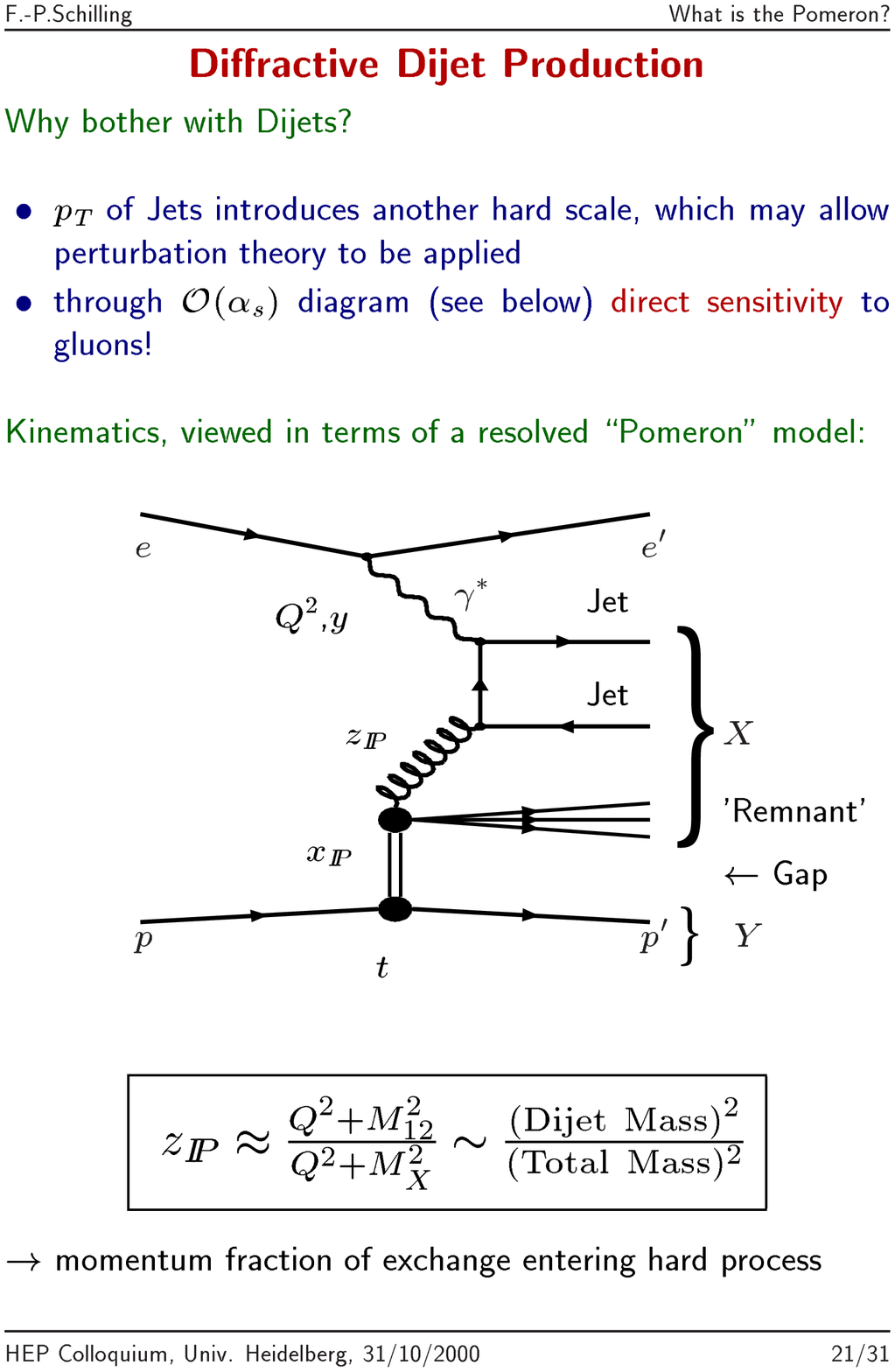,width=0.42\textwidth,clip=}
  \caption{Illustration of diffractive kinematics. Figure taken from
    \cite{SchillingTalk}.}
  \label{fig:Diffraction}
\end{floatingfigure}
While factorisation has long been established in inclusive processes
\cite{GenFact} it has been realised in the past few years
\cite{DiffFact} that it should also hold in more exclusive cases ---
in particular for diffraction, in terms of diffractive parton
distributions $f_{a/p}^\mathrm{diff}(x,x_\pom,\mu^2,t)$, which can be
interpreted loosely as being related to the probability of finding a
parton $a$ at scale $\mu^2$ with longitudinal momentum fraction $x$,
inside a diffractively scattered proton $p$, which in the scattering
exchanges a squared momentum $t$ and loses a longitudinal momentum
fraction $x_\pom$. These kinematic variables are illustrated in
fig.~\ref{fig:Diffraction}.

The dependence of the diffractive parton distributions on so many
variables means that without a large kinematical range
(separately in $x$, $x_\pom$ and $Q^2$, while perhaps integrating over
$t$) it is a priori difficult to thoroughly test diffractive
factorisation. An interesting simplifying assumption is that of Regge
factorisation, where one writes \cite{IngelmanSchlein}
\begin{equation}
  f_{a/p}^\mathrm{diff}(x,x_\pom,\mu^2,t) = |\beta_p(t)|^2
  x_\pom^{-2\alpha(t)}  f_{a/\pom}(x/x_\pom, \mu^2, t)
\end{equation}
the interpretation of diffraction being due to (uncut) pomeron
exchange (first two factors), with the virtual photon probing the
parton distribution of the pomeron (last factor).

As yet no formal justification exists for this extra Regge factorisation.
Furthermore given that diffraction is arguably related to saturation
and high parton densities (assuming the AGK cutting rules
\cite{AGK}) one could even question the validity of arguments
for general diffractive factorisation, which rely on parton densities
being low (as does normal inclusive factorisation).

The experimental study of factorisation in diffraction relied
until recently exclusively on inclusive $F_2^d$ measurements. This was
somewhat unsatisfactory because of the wide range of alternative models
able to reproduce the data and even the existence of significantly
different forms for the $f_{a/\pom}(x/x_\pom, \mu^2, t)$ which gave a
satisfactory description of the data within the Regge factorisation
picture.

\begin{figure}[htbp]
  \begin{center}
    \epsfig{file=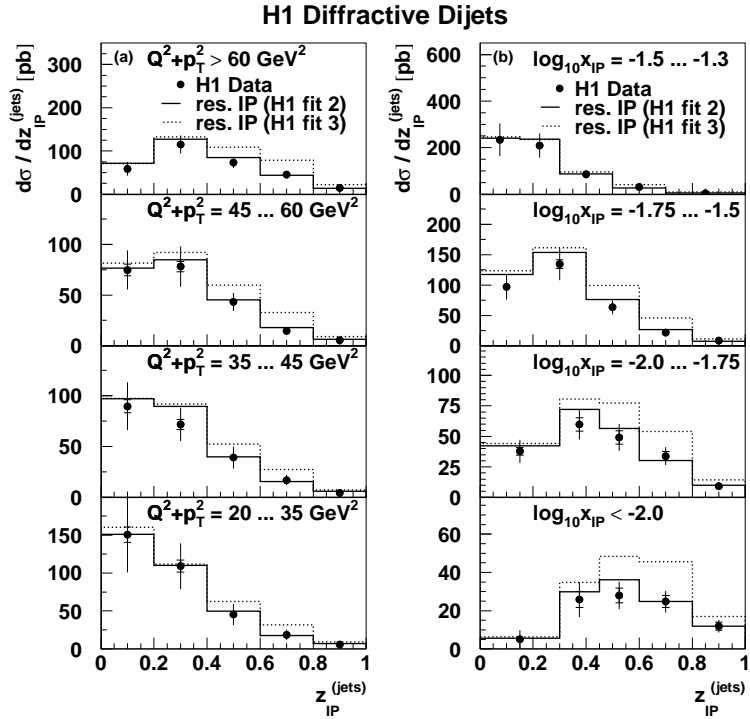,width=0.8\textwidth}
    \caption{Comparisons of H1 diffractive dijet cross sections with
      predictions obtained using the assumption of Regge factorisation
      \cite{SchillingThesis}.}
    \label{fig:DiffDijets}
  \end{center}
\end{figure}
However diffractive factorisation allows one to predict not only
inclusive cross sections but also jet cross sections. Results in the
Regge factorisation framework are compared to data in
figure~\ref{fig:DiffDijets} (taken from \cite{SchillingThesis}),
showing remarkable agreement between the data and the predictions
(based on one of the pomeron PDF fits obtained from $F_2^d$). On the
other hand, when one considers certain other models that work well for
$F_2^d$ the disagreement is dramatic, as for example is shown with the
soft colour neutralisation models \cite{SCI,BGH} in
figure~\ref{fig:DiffSCI}.

\begin{floatingfigure}[r]{0.42\textwidth}
  \mbox{ } \hspace{-0.07\textwidth}
  %\epsfig{file=schilling-feyn-diff.grey.eps,width=0.36\textwidth,clip=}\\
  %\mbox{ } \hspace{-0.05\textwidth}
  \epsfig{file=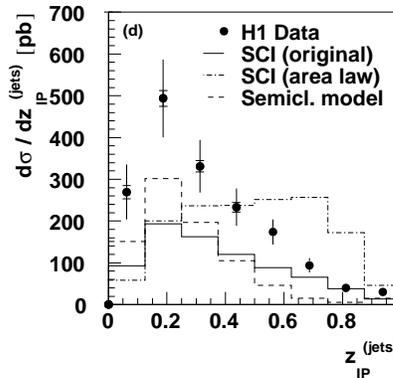,width=0.42\textwidth,clip=}
  \caption{A comparison between diffractive dijet data and results
    from the soft-colour interaction (SCI) \cite{SCI} and
    semiclassical \cite{BGH} models. Figure adapted from
    \cite{SchillingThesis}.}
  \label{fig:DiffSCI}
\end{floatingfigure}
Despite this apparently strong confirmation of diffractive
factorisation, a word of warning is perhaps needed. Firstly there
exist other models which have not been ruled out (for example the
dipole model \cite{DiffDipole}). In these cases it would be of
interest to establish whether these models can be expressed in a way
which satisfies some effective kind of factorisation.

Other important provisos are that a diffractive PDF fit based on more
recent $F_2^{d}$ data has a lower gluon distribution and so leads to
diffractive dijet predictions which are a bit lower than the data,
though still compatible to within experimental and theoretical
uncertainties \cite{DijetTalks}. And secondly that the predictions
themselves are based on the Rapgap event generator \cite{Rapgap} which
incorporates only leading order dijet production.  It would be of
interest (and assuming that the results depend little on the treatment
of the `pomeron remnant,' technically not at all difficult) to
calculate diffractive dijet production to NLO with programs such as
Disent \cite{disent} or Disaster++ \cite{disaster}, using event
generators only for the modelling of hadronisation correction, as is
done in inclusive jet studies.

%----------------------------------------------------------------------
\section{Hadronisation}
\label{sec:Hadr}

Another subject that has seen considerable experimental and
theoretical progress recent years is that of hadronisation. Even at
the relatively high scattering energies involved at LEP and the
Tevatron, for many final state observables non-perturbative
contributions associated with hadronisation are of the same order of
magnitude as next-to-leading order perturbative contributions and
cannot be neglected. With the advent of NNLO calculations in the
foreseeable future the need for a good understanding of hadronisation
becomes ever more important.

Until a few years ago, the only way of estimating hadronisation
corrections in final-state measurements was by comparing the parton
and hadron levels of Monte Carlo event generators. Such a procedure
suffers from a number of drawbacks. In particular the separation
between perturbative and non-perturbative contributions is
ill-defined: for example event generators adopt a prescription for the
parton level based on a cutoff; on the other hand, in fixed-order
perturbative calculations no cutoff is present, and the perturbative
integrals are naively extended into the non-perturbative region ---
furthermore the `illegally-perturbative' contribution associated with
this region differs order by order (and depends also on the
renormalisation scale). 

Additionally, hadronisation corrections obtained from event generators
suffer from a lack of transparency: the hadronisation models are
generally quite sophisticated, involving many parameters, and the
relation between these parameters and the hadronisation corrections is
rarely straightforward.

In the mid 1990's a number of groups started examining approaches for
estimating hadronisation corrections based on the perturbative
estimates of observables' sensitivity to the infrared. This leads to
predictions of non-perturbative corrections which are suppressed by
powers of $1/Q$ relative to the perturbative contribution (for a
review see \cite{BenekeReview}). One of the most successful
applications of these ideas has been to event shapes, for which (in
the formalism of Dokshitzer and Webber \cite{DokWeb})
\begin{equation}
  \langle \cV_\NP \rangle = \langle \cV_\PT \rangle + c_\cV \cP\,,
  \qquad
%  \cP \equiv  \frac{4C_F}{\pi^2}\cM \frac{\mu_I}{Q}
  \cP \equiv  \frac{2C_F}{\pi}\frac{\mu_I}{Q}
\left\{ \alpha_0(\mu_I)- \as(Q)
  - \order{\as^2} \right\}\>,
\end{equation}%
\begin{floatingfigure}[r]{0.42\textwidth}
  \mbox{ } \hspace{-0.09\textwidth}
  \epsfig{file=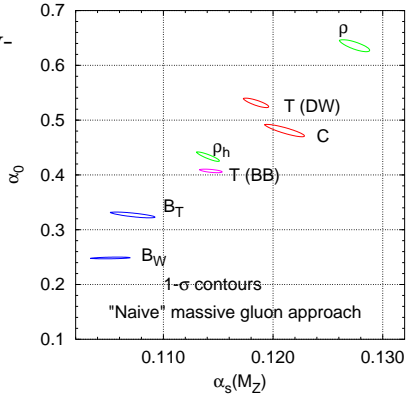,width=0.42\textwidth}
  \caption{Results of fits to $\ee$ mean event shapes using original,
    `naive' calculations for $c_\cV$.}
  \label{fig:contnaive}
\end{floatingfigure}%
\noindent where $c_\cV$ is a perturbatively calculable
observable-dependent coefficient and $\cP$ governs the size of the
power correction.  The quantity $\alpha_0(\mu_I)$, which can be
interpreted as the mean value of an infrared finite effective coupling
in the infrared (up to an infrared matching scale $\mu_I$,
conventionally chosen to be $2$~GeV), is hypothesised to be
\emph{universal}. The terms in powers of $\as$ are subtractions of
pieces already included in the perturbative prediction for the
observable.

It is interesting to see the progress that has been made in our
understanding of these effects. The first predictions for the $c_\cV$
coefficients were based on calculations involving the Born
configuration plus a single `massive' (virtual) gluon. Fitting
$\alpha_0$ and $\as$ to data for mean values of $\ee$ event-shapes,
using the original predictions for the $c_\cV$, leads to the results
shown in figure~\ref{fig:contnaive}.

At the time of the original predictions, however, much of the data
used to generate fig.~\ref{fig:contnaive} was not yet in existence
(which is perhaps fortunate --- had fig.~\ref{fig:contnaive} been
around in 1995, the field of $1/Q$ hadronisation corrections might not
have made it past early childhood). Rather, various theoretical
objections (\eg \cite{NasonSeymour}) and the gradual appearance of new
data, especially for the broadenings, forced people to refine their
ideas. 

\begin{floatingfigure}[r]{0.42\textwidth}
  \mbox{ } \hspace{-0.09\textwidth}
  \epsfig{file=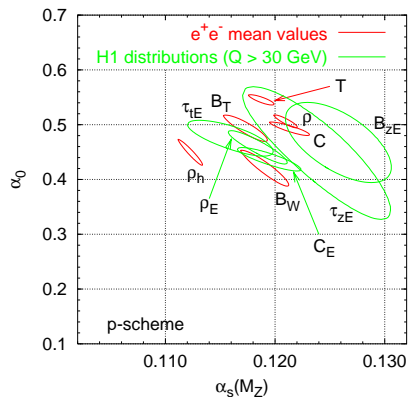,width=0.42\textwidth}
  \caption{Results of fits to $\ee$ mean event shapes and DIS
    distributions with state-of-the-art account of NP contributions.}
  \label{fig:cont2002}
\end{floatingfigure}%
Among the developments was the realisation that to control the
normalisation of the $c_\cV$ it is necessary to take into account the
decay of the massive, virtual, gluon (the reason for the two thrust
results in fig.~\ref{fig:contnaive} was the existence of two different
conventions for dealing with the undecayed massive gluon)
\cite{MilanFactor}. It was also realised that it is insufficient to
consider a lone `non-perturbative' gluon, but rather that such a gluon
must be taken in the context of the full structure of soft and
collinear perturbative gluon radiation \cite{ResummedNP}.  Another
discovery was that hadron-masses can be associated with universality
breaking $1/Q$ power corrections in certain definitions of observables
\cite{SalamWicke} and when testing the universality picture all
observables should be measured in an appropriate common `hadron-mass'
scheme.

Results incorporating these theoretical developments are shown in
figure~\ref{fig:cont2002}. As well as $\ee$ mean event shapes we also
include recent results using resummed DIS event shapes
\cite{DasguptaSalam}, fitted to H1 distributions \cite{H1dist}. The
agreement between observables, even in different processes, is
remarkable, especially compared to fig.~\ref{fig:contnaive}, and a
strong confirmation of the universality hypothesis.\footnote{It should
  be noted that results for certain $\ee$ distributions
  \cite{KluthFits} and DIS means \cite{H1dist,ZEUSmeans} are not quite
  as consistent. Though this remains to be understood, it may in part
  be associated with the particular fit ranges that are used.}

This is not to say that the field has reached maturity. In the above
fits the approximation has been made that non-perturbative corrections
just shift the perturbative distribution \cite{DokWebShift}, however
there exists a considerable amount of recent work which examines the
problem with the more sophisticated `shape-functions' approach
\cite{ShapeFunctions} in particular in the context of the Dressed
Gluon Exponentiation approximation \cite{DGE}.  An important point
also is that all the detailed experimental tests so far are for 2-jet
event shapes, where there exists a solid theoretical justification
based on the Feynman tube model \cite{FeynmanTube}, \ie longitudinal
boost invariance. It will be of interest to see what happens in
multi-jet tests of $1/Q$ hadronisation corrections where one
introduces both non-trivial geometry and the presence of gluons in the
Born configuration \cite{MilanMultiJet}. Finally we note the
provocative analysis by the Delphi collaboration \cite{Delphi} where
they show that a renormalisation-group based fit prefers an absence of
hadronisation corrections, at least for mean values of event shapes,
as well as leading to highly consistent values for $\as$ across a
range of event-shapes.

%----------------------------------------------------------------------
\section{Heavy quark ($b$) production}
\label{sec:Heavy}

For light quarks (and gluon) it is impossible to make purely
perturbative predictions of their multiplicity or of their
fragmentation functions because of soft and collinear divergences. For
heavy quarks however, these divergences are cut off by the quark mass
itself, opening the way to a range of perturbative predictions and
corresponding tests of QCD.

\begin{figure}[htbp]
  \begin{center}
    \begin{minipage}{0.50\textwidth}
      \epsfig{file=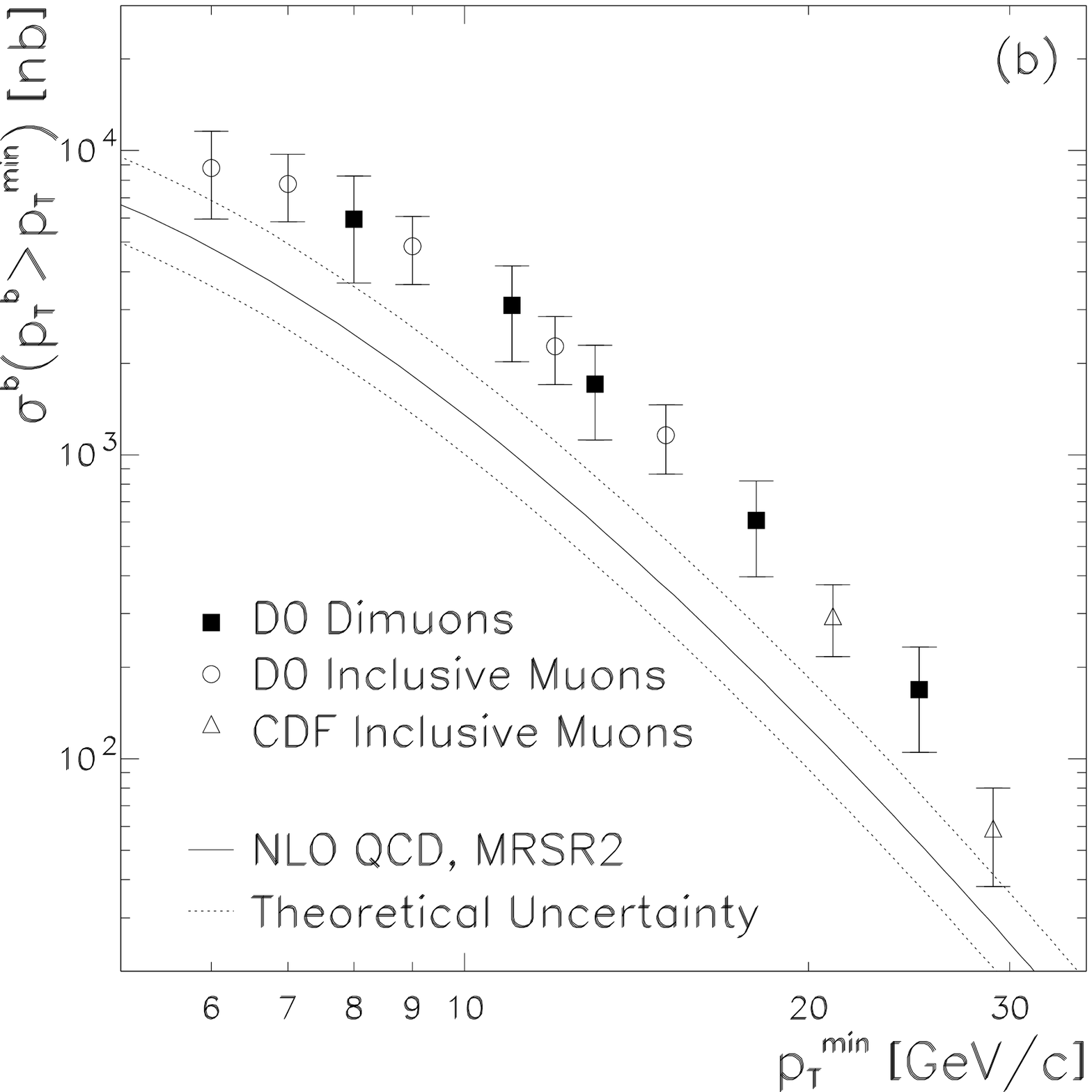,width=\textwidth}
    \end{minipage}\hfill
    \begin{minipage}{0.48\textwidth}
      \mbox{ } \vspace{-0.03\textwidth}\\
      \epsfig{file=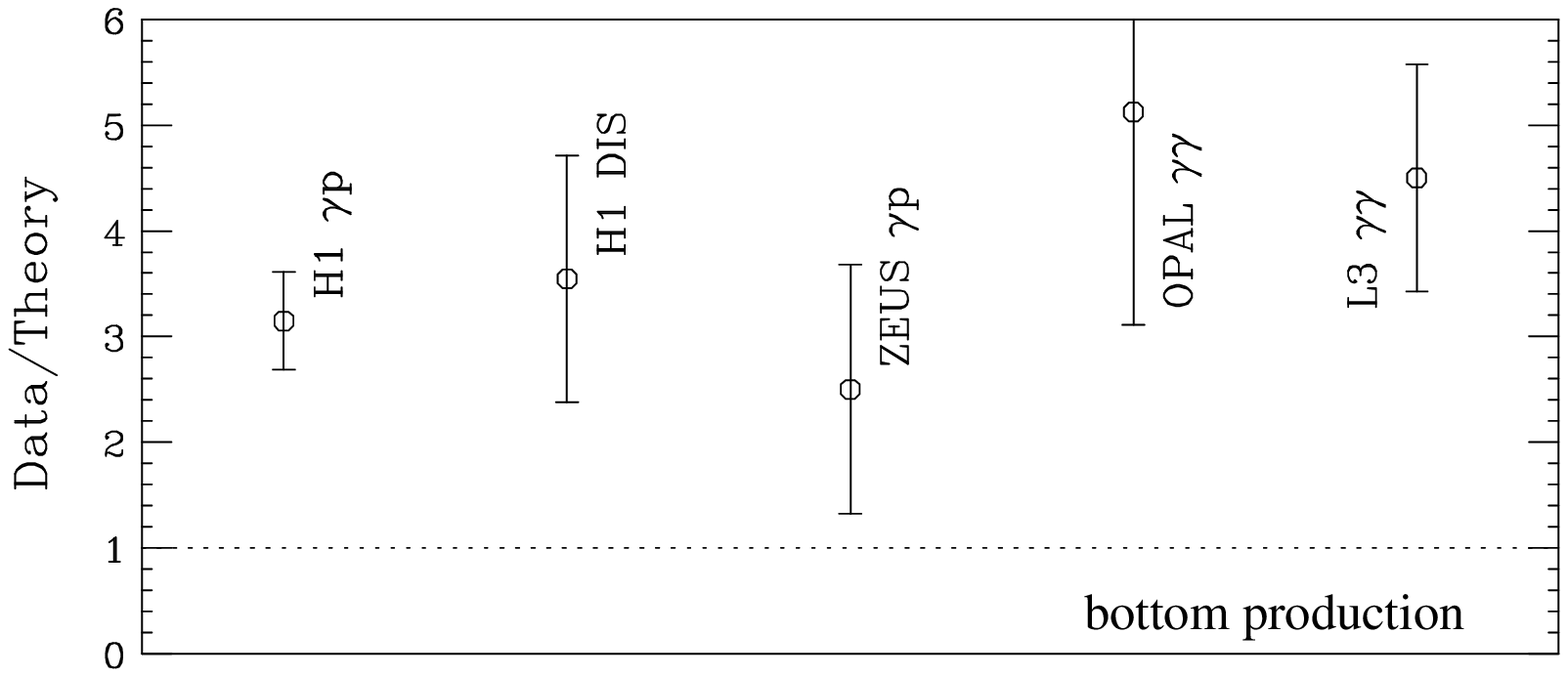,width=\textwidth}%
\vspace{0.05\textwidth}\\ 
      \epsfig{file=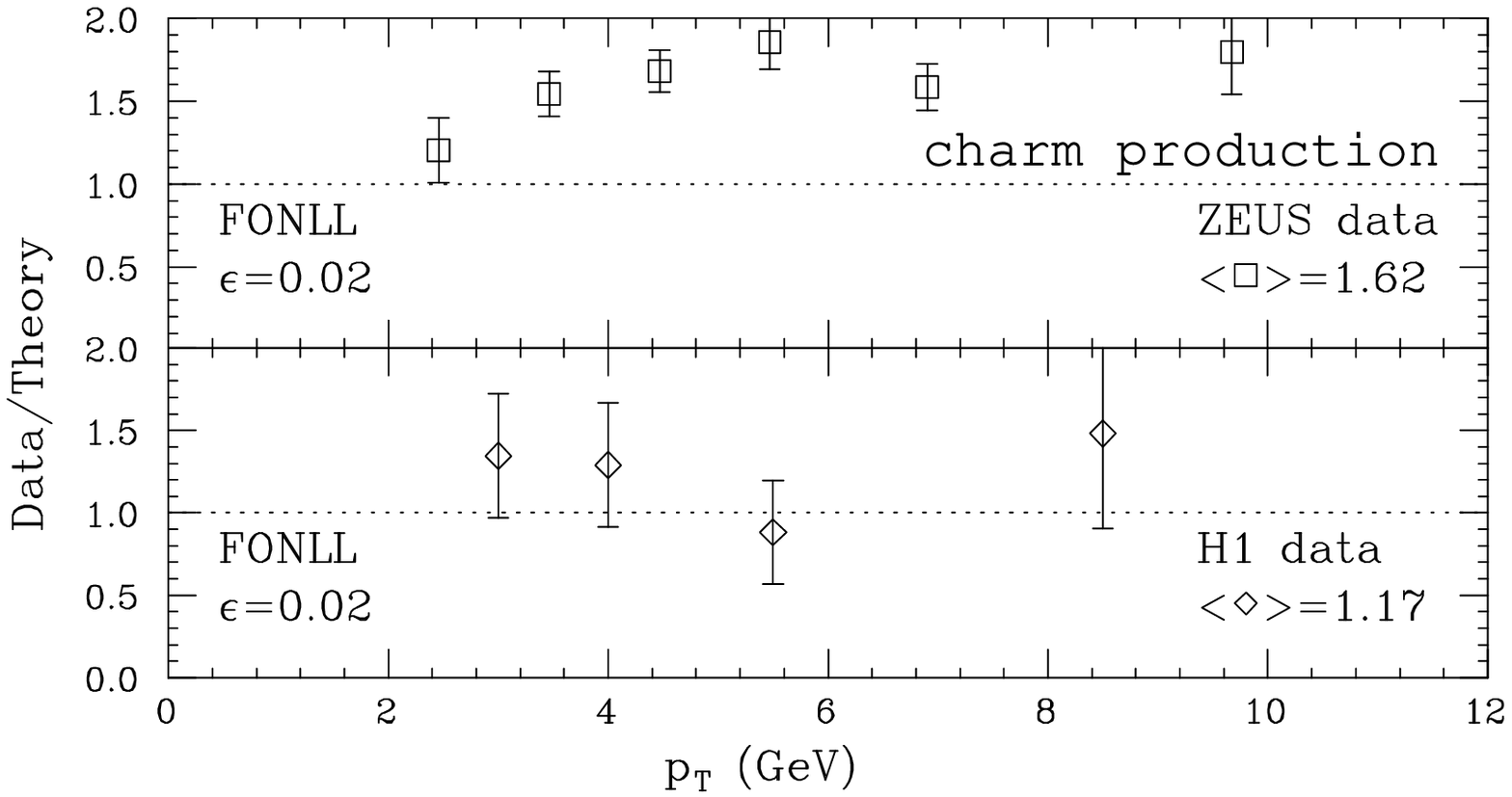,width=\textwidth}
    \end{minipage}
    \caption{Left: $b$-quark $p_t$ distribution at the Tevatron
      \cite{bTevatron}; upper right: summary of open $b$ cross
      sections in $\gamma p$, DIS and $\gamma \gamma$ collisions,
      normalised to theoretical expectations (figure taken from
      \cite{bcFrixione}); lower right: ratio of experiment to theory
      for the charm $p_t$ distribution at HERA (taken from
      \cite{bcFrixione}).}
    \label{fig:bcSummary}
  \end{center}
\end{figure}

It is therefore particularly embarrassing that there should be a
significant discrepancy in most experiments (but not all, \eg
\cite{HeraB}) where the QCD bottom production cross section has been
measured. The situation is shown in figure~\ref{fig:bcSummary} for
Tevatron, HERA and LEP results, illustrating the systematic excess of
a factor of three between measurements and NLO calculations. To add to
the puzzle, the agreement for charm production (which if anything
should be worse described because of the smaller mass) is considerably
better across a range of experiments (see \eg the lower-right plot of
fig.~\ref{fig:bcSummary}).

Aside from the intrinsic interest of having a good understanding of
$b$-production in QCD, one should keep in mind that $b$-quarks are
widely relied upon as signals of Higgs production and in searches for
physics beyond the standard model, so one needs to have confidence in
predictions of the QCD background.

We shall discuss a couple of explanations that have been proposed for
the excess at the Tevatron (the excesses in other experiments are more
recent and have yet to be addressed in the same detail). Indeed, one
hypothesis is precisely that we are seeing a signal of light(ish)
gluino production. Another is that bottom fragmentation effects have
been incorrectly accounted for. A third explanation, discussed in
detail in another of the opening plenary talks \cite{deRoeckTalk} is
associated with unintegrated $k_t$ distributions and small-$x$
resummations.

%......................................................................
\subsection{The SUSY hypothesis}
\begin{floatingfigure}[r]{0.42\textwidth}
  \mbox{ } \hspace{-0.06\textwidth}
  \epsfig{file=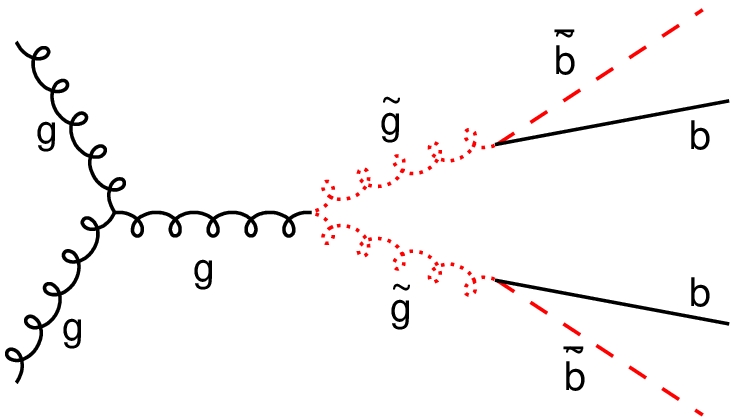,width=0.35\textwidth}\\
  \mbox{ } \hspace{-0.04\textwidth}
  \epsfig{file=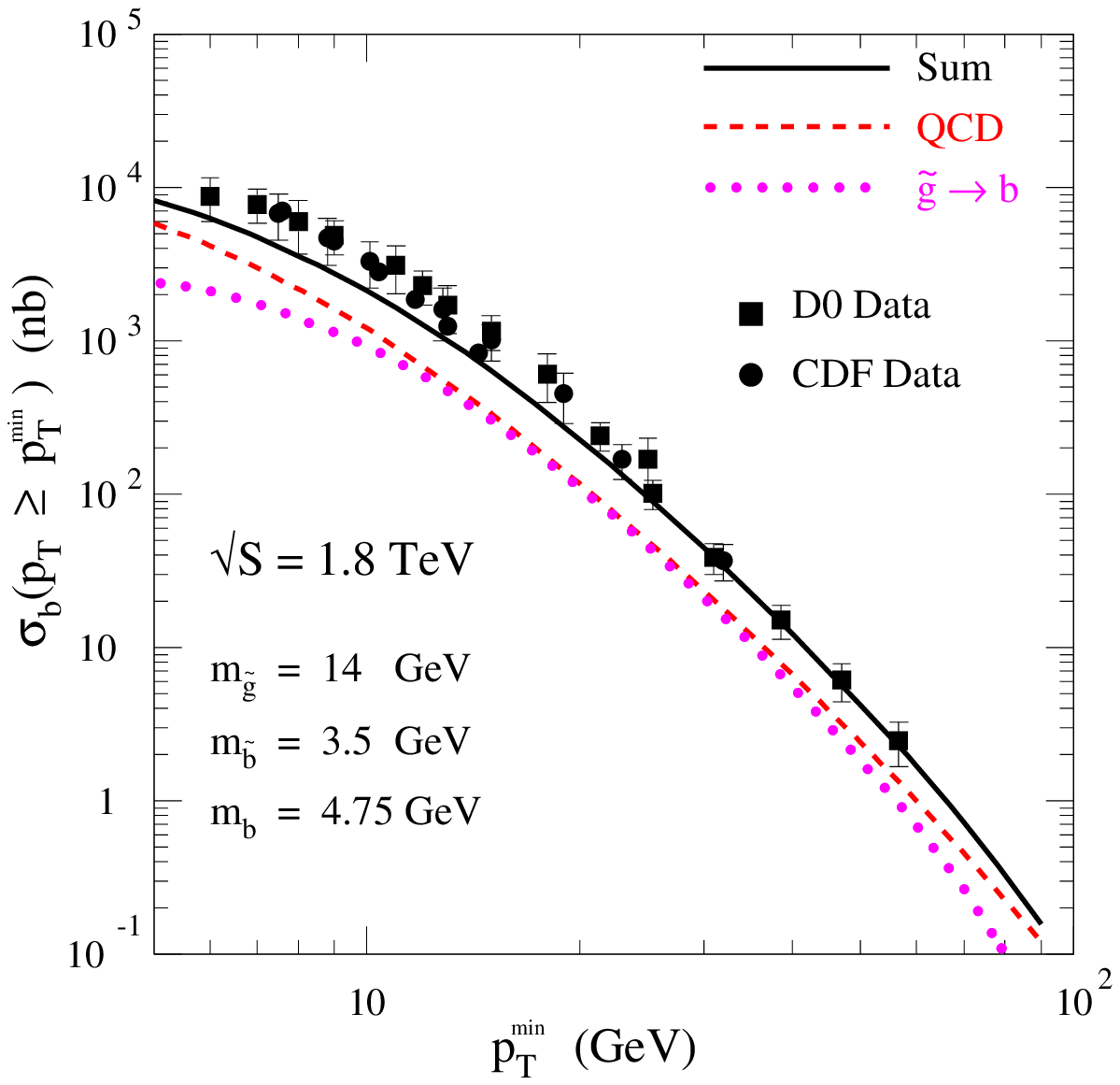,width=0.42\textwidth}
  \caption{Above: a potential SUSY contribution to the $b$-quark
    excess. Below: the resulting modification of the bottom $p_t$
    distribution at the Tevatron \cite{Berger}.}
  \label{fig:gluinos}
\end{floatingfigure}
In \cite{Berger} it has been argued that a possible explanation of the
Tevatron $b$-quark excess is the production of a pair of light gluinos
with a mass of order $14$~GeV which then decay to sbottoms ($\sim
3.5$~GeV) and bottoms, as in fig.~\ref{fig:gluinos}. The mixing angles
are chosen such that the sbottom decouples from the $Z$ at LEP,
accounting for its non-observation there.

At moderate and larger $p_t$, the contribution from this process is
about as large as that from NLO QCD and so it brings the overall
production rate into agreement with the data.

There are a number of other consequences of such a scenario: one is
the production of like-sign $b$ quarks (as in the Feynman graph of
fig.~\ref{fig:gluinos}), which could in principle be observed at the
Tevatron, although it would need to be disentangled from $B_0$-${\bar
  B}_0$ mixing. Another is that the running of $\as$ would be modified
significantly above the gluino mass, leading to an increase of about
$0.007$ in the running to $M_Z$ of low $Q$ measurements of $\as$. This
seems to be neither favoured nor totally excluded by current $\as$
measurements.

Though they have not provided a detailed analysis, the authors of
\cite{Berger} also consider the implications for HERA. There it seems
that the enhancement of the $b$-production rates is too small to
explain the data (because of the suppression due to the gluino mass).

%......................................................................
\subsection{The fragmentation explanation}

In any situation where one sees a significant discrepancy from QCD
expectations it is worth reexamining the elements that have gone into
the theoretical calculation. Various groups have considered issues
related to $b$ fragmentation and found significant effects, which
could be of relevance to the Tevatron results (see for example
\cite{OldFragB}). However a recent article by Cacciari and Nason
\cite{CaccNas} is particularly interesting in that it makes use of the
full range of available theoretical tools to carry out a unified
analysis all the way from the $\ee$ data, used to constrain the
$b$-quark fragmentation function, through to expectations for the
Tevatron. It raises a number of important points along the
way.\footnote{The reader is referred to their article for full
  references to the `ingredients' used at different stages of the
  analysis.}

To be able to follow their analysis it is worth recalling how one
calculates expectations for processes involving heavy quarks.  The
cross section for producing a $b$-quark with a given $p_t$ (or even
integrated over all $p_t$) is finite, unlike that for a light quark.
This is because the quark mass regulates (cuts-off) the infrared
collinear and soft divergences which lead to infinities for massless
quark production. But infrared finiteness does not mean infrared
insensitivity and to obtain a $B$-meson $p_t$ distribution from a
$b$-quark ${\hat p}_t$ distribution, one needs to convolute with a
fragmentation function,
\begin{equation}
  \label{eq:frag}
 \displaystyle \underbrace{\frac{d\sigma}{dp_t}}_{\mathrm{
    measured,\;e.g.\;}B_0} = \int d \hat{p}_t dz
  \underbrace{\frac{d\sigma}{d\hat{p}_t}}_{\mathrm{ PT\,
      QCD,}\;b\;\mathrm{quark}}
  \underbrace{D(z)}_{\mathrm{ fragmentation}} \delta(p_t - z \hat{p}_t) \,.
\end{equation}
The details of the infrared finiteness of the $b$-quark production are
such that $\langle z D(z) \rangle$ is $1 - \order{\Lambda/m_b}$, where
the origin of the $\Lambda/m_b$ piece is closely related to that of
the $\Lambda/Q$ power corrections discussed in the previous section
\cite{NasonWebber}.

There are various well-known points to bear in mind about
fragmentation functions.  Firstly, in close analogy to the
hadronisation corrections discussed earlier (and of course structure
functions), the exact form for the fragmentation function will depend
on the perturbative order at which we define eq.~\eqref{eq:frag}.
Secondly, while for $p_t \sim m_b$ we are free to use fixed order (FO)
perturbative predictions, for $p_t \gg m_b$ there are large
logarithmically enhanced terms, which need to be resummed.  The
technology for doing this currently exists to next-to-leading
logarithmic (NLL) order. In the intermediate region $p_t \gtrsim m_b$
the two approaches can be combined to give FONLL predictions
\cite{FONLLb,OleariNason} (strictly this can be used even for $p_t \gg
m_b$).

Having established these points we can consider what has been done by
Cacciari and Nason \cite{CaccNas}. Firstly they discuss moments of the
fragmentation function $\langle z^{N-1} D(z)\rangle$. This is because
for a steeply falling perturbative ${\hat p}_t$ distribution in
eq.~\eqref{eq:frag}, $\frac{d\sigma}{d\hat{p}_t} \sim 1/{\hat p}_t^N$,
after integrating out the $\delta$-function to give ${\hat p}_t =
p_t/z$, one obtains the result 
\begin{equation}
  \left.\frac{d\sigma}{dp_t}\right|_{B\mathrm{-meson}} = 
\langle z^{N-1} D(z)\rangle  \left.
\frac{d\sigma}{d\hat{p}_t}\right|_{b\mathrm{-quark}} \,,
\end{equation}
where for the Tevatron $N\simeq5$.

\begin{floatingfigure}[r]{0.47\textwidth}
  \mbox{ } \hspace{-0.07\textwidth}
  \epsfig{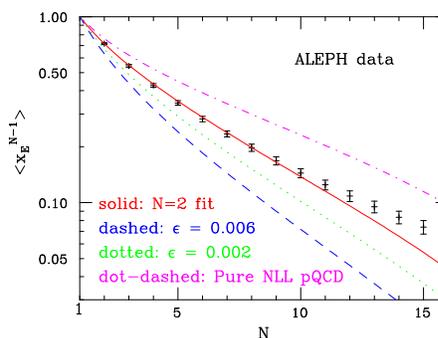}\\
  \caption{Moments of the momentum fraction carried by $B$-mesons in
    $\ee$, compared to NLL predictions with and without fragmentation
    functions \cite{CaccNas}.}
  \label{fig:eefrag}
\end{floatingfigure}
The cleanest place to constrain $b$ fragmentation is in $\ee$
collisions.  Figure~\ref{fig:eefrag} shows moments of the momentum
fraction (with respect to $Q/2$) carried by $B$-mesons as measured by
Aleph \cite{AlephBFrag}. The (magenta) dot-dashed curve shows the
purely perturbative NLL prediction, which is clearly above the data.
The dashed curve shows what happens when one includes the convolution
with an $\epsilon=0.006$ Peterson fragmentation function
\cite{Peterson}. Why this particular function? Simply because it is
the one included in certain Monte Carlo event generators and used
widely by experimental collaborations that have compared measured and
theoretical $p_t$ distributions. The data point for the $N=5$ moment
is $50\%$ higher than the theoretical expectation with this
fragmentation function.

Of course we don't expect agreement: the $\epsilon=0.006$ Peterson is
widely used in Monte Carlos where one has only leading-logs. But we
are interested in NLL calculations and the fragmentation function
needs to be refitted. The authors of \cite{CaccNas} take the
functional form of \cite{CaccNasFragFun}, fitted to the $N=2$ moment,
to give the solid curve.

%A good form at NLL order turns out to be the
%fragmentation function of \cite{CaccNasFragFun}, which when fitted to
%the $N=2$ moment, gives the solid curve.

\begin{figure}[t]
  \begin{center}
    \begin{minipage}{0.49\textwidth}
      \epsfig{file=CaccNas-B-hadrons.eps,width=\textwidth}
    \end{minipage}\hfill
    \begin{minipage}{0.49\textwidth}
      \epsfig{file=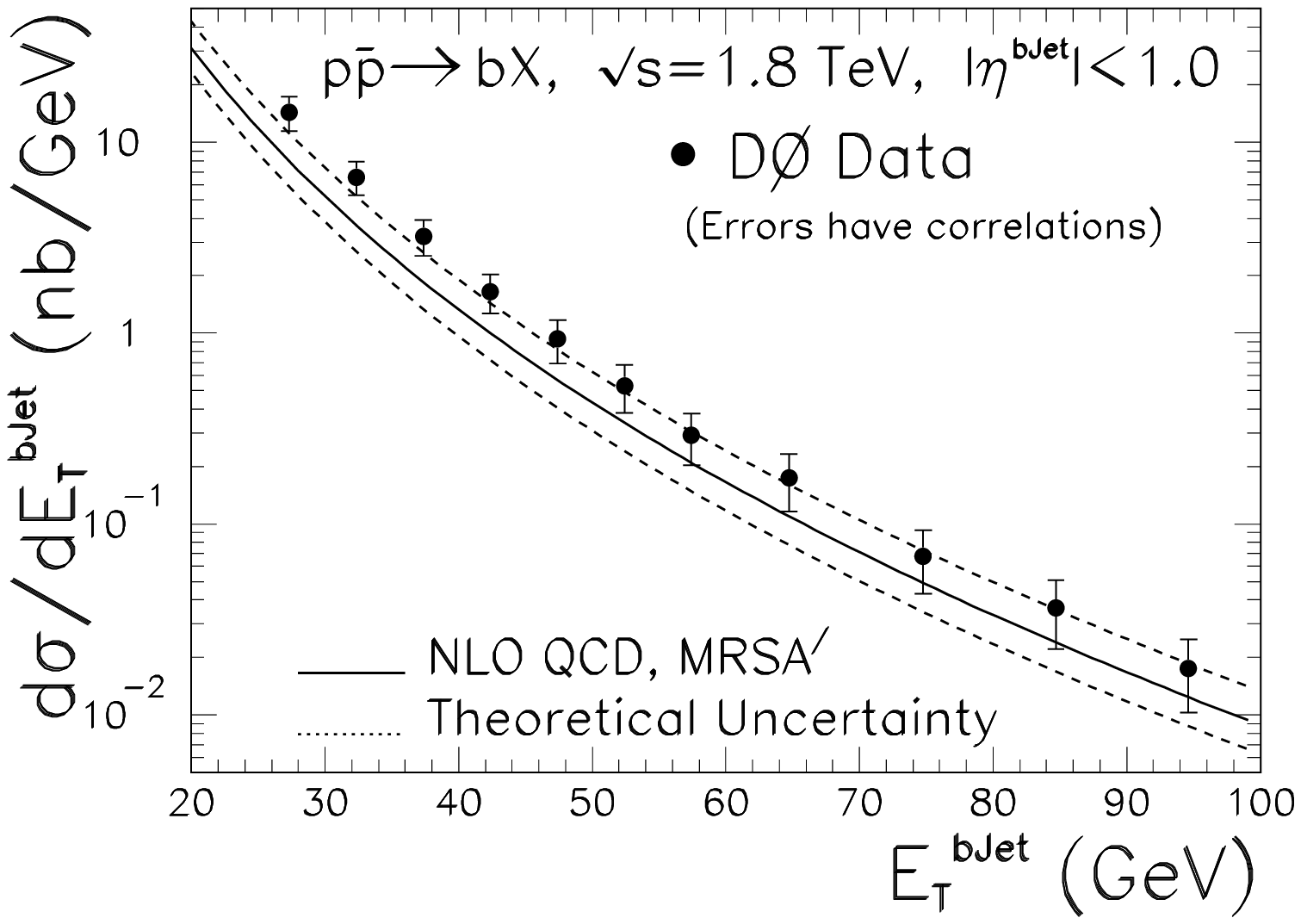,width=\textwidth}
    \end{minipage}
    \caption{Left: result for $B$-meson production \cite{CDFhadrB}
      compared to the FONLL prediction with the `$N=2$' fragmentation
      function \cite{CaccNas}. Right: results for $b$-jet production
      \cite{D0bjet} compared to the NLO predictions
      \cite{FrixioneMangano}.}
    \label{fig:GoodB}
  \end{center}
\end{figure}

The next step in the Cacciari and Nason analysis should simply have
been to take the FONLL calculation of bottom production at the
Tevatron \cite{FONLLb}, convolute with their new fragmentation
function and then compare to data. This however turns out to be
impossible for most of the data, because it has already been
deconvoluted to `parton-level' (in some cases with the
$\epsilon=0.006$ Peterson fragmentation function). So they are only
able to compare with the recent CDF data \cite{CDFhadrB} for
$B$-mesons, shown in the left-hand plot of figure \ref{fig:GoodB}. The
dashed curve is the central result, while the solid ones are those
obtained when varying the factorisation and renormalisation scales by
a factor of two.\footnote{A point worth keeping in mind
  \cite{DokPrivComm} is that the central scale choice $\mu =
  \sqrt{p_t^2 + m_b^2}$ is not universally accepted as being optimal
  --- indeed for $p_t \gtrsim m_b$, a scale choice of $\mu = p_t$ is
  equally justifiable, and would have a non-negligible effect on the
  predictions.}  The dotted curve shows the results that would have
been obtained with the Peterson fragmentation function.  Predictions
with FO (generally used in previous comparisons) rather than FONLL
would have have been $20\%$ lower still.

Another interesting approach to the problem is to eliminate the
fragmentation aspects altogether, which can be achieved by looking at
the $E_t$ distribution of $b$-jets, without specifically looking at
the $b$ momentum \cite{FrixioneMangano}. This has been examined by the
D0 collaboration \cite{D0bjet} and the comparison to NLO predictions
is shown in the right-hand plot of figure~\ref{fig:GoodB}. Though in a
slightly different $E_t$ range, the relation between theory and data
is similar to that in the Cacciari-Nason approach for $B$-mesons:
there is a slight excess in the data but not significant compared to
the uncertainties. A minor point to note in the study of $b$-jets is
that there are contributions $\as^n \ln^{2n-1} E_t/m_b$ from soft and
collinear logs in the multiplicity of gluons which can then branch
collinearly to $b{\bar b}$ pairs \cite{FrixioneManganoRef12}. At very
large $E_t$ these terms would need to be resummed.

So overall, once one has a proper theoretical treatment, including
both an appropriate fragmentation function and, where relevant, an
FONLL perturbative calculation, it is probably fair to say that the
excess of $b$-production at the Tevatron is not sufficiently
significant to be worrisome (or evidence for supersymmetry).

At some of the other experiments where an excess of $b$-production is
observed a number of the same issues arise, in particular relative to
the use of the $\epsilon = 0.006$ Peterson fragmentation function and
the presentation of results at parton level rather than hadron
level. However fragmentation is less likely to be able to explain the
discrepancies, because of the lower $p_t$ range. 

%----------------------------------------------------------------------
\section{Event generators at NLO}
\label{sec:EvGen}

The problem of matching event generators with fixed order calculations
is one of the most theoretically active areas of QCD currently, and
considerable progress has been made in the past couple of years. This
class of problems is both of intrinsic theoretical interest in that it
requires a deep understanding of the structure of divergences in QCD
and of phenomenological importance because of the need for accurate
and reliable Monte Carlo predictions at current and future colliders.

Two main directions are being followed: one is the matching of
event-generators with leading-order calculations of $n$-jet production
(where $n$ may be relatively high), which is of particular importance
for correctly estimating backgrounds for new-particle searches
involving cascades of decays with many resulting jets. For a
discussion of this subject we refer the reader to the contributions to
the parallel sessions \cite{PrllLonnbladKrauss}. 

The second direction, still in its infancy, is the matching of event
generators with next-to-leading order calculations (currently
restricted to low numbers of jets), which is necessary for a variety
of purposes, among them the inclusion of correct rate estimates
together with consistent final states, for processes with large NLO
corrections to the Born cross sections (\eg $K$ factors in $pp$ and
$\gamma p$ collisions, boson-gluon fusion at small-$x$ in DIS).

While there have been a number of proposals concerning NLO matching,
many of them remain at a somewhat abstract level. We shall here
concentrate on two approaches that have reached the implementational
stage. As a first step, it is useful to recall why it is non-trivial
to implement NLO corrections in an event generator. Let us use the
toy model introduced by Frixione and Webber \cite{FrixioneWebber},
involving the emission only of `photons' (simplified, whose only
degree of freedom will be their energy) from (say) a quark whose
initial energy is taken to be $1$. For a system which has radiated $n$
photons we write a given observable as $O(E_q, E_{\gamma_1},\ldots,
E_{\gamma_n})$. So for example at the Born level, the observable has
value $O(1)$. At NLO we have to integrate over the momentum of an
emitted photon, giving the following contribution to the mean value of
the observable:
\begin{equation}
  \label{eq:real}
  \alpha \int_0^1 \frac{dx}{x}\, R(x)\, O(1-x, x)\,,
\end{equation}
where $R(x)$ is a function associated with the real matrix element for
one-photon emission. There will also be NLO virtual corrections and
their contribution will be
\begin{equation}
  \label{eq:virt}
  -\alpha \, O(1) \int_0^1 \frac{dx}{x}\, V(x)\,,
\end{equation}
where $V(x)$ is related to the matrix element for virtual
corrections. 

The structure of $dx/x$ divergences is typical of field theory.
Finiteness of the overall cross section implies that for $x\to0$,
$R(x) = V(x)$. This means that for an infrared safe observable (\ie
one that satisfies $\lim_{x\to 0} O(1-x,x) = O(1)$), the
$\order{\alpha}$ contribution to the mean value of the observable is
also finite.  However any straightforward attempt to implement
eqs.~\eqref{eq:real} and \eqref{eq:virt} directly into an event
generator will lead to problems because of the poor convergence
properties of the cancellation between divergent positively and
negatively weighted events corresponding to the real and virtual
pieces respectively. So a significant part of the literature on
matching NLO calculations with event generators has addressed question
of how to recast these divergent integrals in a form which is
practical for use in an event generator (which must have good
convergence properties, especially if each event is subsequently going
to be run through a detector simulation). The second part of the
problem is to ensure that the normal Monte Carlo event generation
(parton showering, hadronisation, etc.) can be interfaced with the NLO
event generation in a consistent manner.

One approach that has reached the implementational stage could be
called a `patching together' of NLO and MC. It was originally proposed
in \cite{BaerReno} and recently further developed in \cite{Potter}
and extended in \cite{Dobbs}. There one chooses a cutoff
$x_\mathrm{zero}$ on the virtual corrections such that the sum of Born
and virtual corrections gives zero:
\begin{equation}
  1 - \alpha \int_{x_\mathrm{zero}}^1 \frac{dx}{x} V(x) \equiv 0\,.
\end{equation}
It is legitimate to sum these two contribution because they have the
same (Born) final state. Then for each event, a real emission of
energy $x$ is generated with the distribution $dx/x R(x)$ and with the
same cutoff as on the virtuals. The NLO total cross section is
guaranteed to be correct by construction:
\begin{equation}
  \sigma_\mathrm{NLO} \equiv \sigma_0 \alpha \int_{x_\mathrm{zero}}^1
  \frac{dx}{x} R(x)\,.
\end{equation}
The next step in the event generation is to take an arbitrary
separation parameter $x_\mathrm{sep}$, satisfying $x_\mathrm{zero} <
x_\mathrm{sep} < 1$. For $x>x_\mathrm{sep}$ the NLO emission is
considered hard and kept (with ideally the generation of normal Monte
Carlo showering below scale $x$, as in the implementation of
\cite{Dobbs}). For $x<x_\mathrm{sep}$ the NLO emission is thrown away
and normal parton showing is allowed below scale $x$.\footnote{For
  simplicity, many important but sometimes tricky technical details
  have been left out. This will also be the case for the merging
  procedure discussed lower down.}

Among the advantages are that the events all have positive and uniform
weights. And while the computation of $x_\mathrm{zero}$ is
non-trivial, the method requires relatively little understanding of
the internals of the event generator (which are often poorly
documented and rather complicated). However the presence of the
separation parameter $x_\mathrm{sep}$ is in principle problematic:
there can be discontinuities in distributions at $x_\mathrm{sep}$,
certain quantities (for example the probability for a quark to have
radiated an amount of energy less than some $x_\mathrm{r}$ which is
below $x_\mathrm{sep}$) will not quite be correct to NLO and above
$x_\mathrm{sep}$ potentially large logarithms of $x_\mathrm{sep}$ are
being neglected.  These last two points mean that for each new
observable that one studies with the Monte Carlo program, one should
carry out an analysis of the $x_\mathrm{sep}$ dependence (varying it
over a considerable range, not just a factor of two as is sometimes
currently done).

A rather different approach (which we refer to as `merging') has been
developed by Frixione and Webber in \cite{FrixioneWebber}.\footnote{A
  number of aspects of the work of Collins and collaborations
  \cite{CollinsMC} may actually be equivalent, though presented in a
  rather different framework. Related issues are discussed also in
  \cite{Mrenna}.} They specify a number of conditions that must be
satisfied by a Monte Carlo at NLO (MC{@}NLO): i) all observables
should be correct to NLO; ii) soft emissions should be treated as in a
normal event generator and hard emissions as in an NLO calculation;
iii) the matching between the hard and soft regions should be smooth.
Their approach exploits the fact that Monte Carlo programs already
contain effective real and virtual NLO corrections,
\begin{equation}
         \pm \alpha \frac{dx}{x} M(x) \quad
      \mbox{for}\;\;\;^\mathrm{real}_\mathrm{virtual}\,.
\end{equation}
Because Monte Carlo programs are designed to correctly reproduce the
structure of soft and collinear divergences, $M(x)$ has\footnote{Or
  rather, `should have.' In practice the divergence structure of
  large-angle soft-gluon emission is not always properly treated in
  event generators, which leads to some extra complications in the
  MC@NLO approach.} the property that for $x\to0$, $M(x) = R(x) =
V(x)$, \ie the divergent part of the NLO corrections is already
included in the event generator. This can be exploited when adjusting
the Monte Carlo to be correct to NLO, because the regions that need
adjusting are the hard regions, but not the (soft) divergent regions.
Specifically the method introduced in \cite{FrixioneWebber} can be
summarised by the formula
\begin{multline}
  I_{\mathrm{MC,Born}} - \alpha\,
  I_{\mathrm{MC,Born}} \int \frac{dx}{x}\left( { V(x)} { -
      M(x)} \right)\\ + \alpha \int \frac{dx}{x}\left( { R(x)}
    { - M(x)} \right)I_{\mathrm{MC,Born}+x}\,.
\end{multline}
$I_{\mathrm{MC,Born}}$ is to be read `interface to Monte Carlo.' It
means that one should generate a Monte Carlo event starting from the
Born configuration (or from the Born configuration plus a photon in
the case of $I_{\mathrm{MC,Born}+x}$). Since at the Born level,
$I_{\mathrm{MC,Born}}$ already contains effective real and virtual
corrections which go as $\pm \alpha M(x)/x$, when evaluating the NLO
corrections to the MC, these pieces should be subtracted from the full
NLO matrix elements. Because $M(x)$ and $R(x)$ (or $V(x)$) have the
same $x\to0$ limit, the real and virtual integrals are now
individually finite and well-behaved, which means that the Monte Carlo
only needs only a small, $\order{\alpha}$, correction in order for it
to be correct to NLO.

\begin{floatingfigure}[r]{0.47\textwidth}
  \mbox{ } \hspace{-0.07\textwidth}
  \epsfig{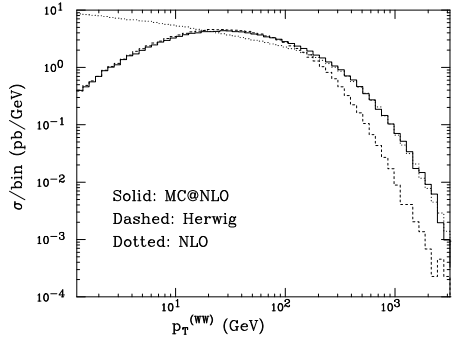}\\
  \caption{Transverse momentum distribution of $W^+W^-$ pairs in $pp$
    collisions calculated at NLO, with Herwig (multiplied a
    $K$-factor) and with MC@NLO \cite{FrixioneWebber}.}
  \label{fig:mcnlo}
\end{floatingfigure}
Illustrative results from this approach are shown in
figure~\ref{fig:mcnlo} for the transverse momentum distribution of a
$W^+W^-$ pair in hadron-hadron collisions. In the low transverse
momentum region (which requires resummation --- the pure NLO
calculation breaks down) MC@NLO clearly coincides with the Herwig
results, while at high transverse momentum it agrees perfectly with
the NLO calculation (default Herwig is far too low).

So this procedure has several advantages: it is a smooth procedure
without cutoffs; the predictions are guaranteed to be correct at NLO
and it does not break the resummation of large logarithms. From a
practical point of view it has the (minor) drawback of some events
with negative weights, however the fraction of negative weight events
is low (about $10\%$ in the example shown above) and they are uniform
negative weights, so they should have little effect on the convergence
of the results. Another limitation is that to implement this method it
is necessary that one understand the Monte Carlo event generator
sufficiently well as to be able to derive the function $M(x)$, \ie the
effective NLO correction already embodied in the event generator.
This however is almost certainly inevitable: there is no way of
ensuring a truly NLO result without taking into account what is
already included in the event generator.

%----------------------------------------------------------------------
\section{Conclusions: testing QCD?}
\label{sec:Concl}

An apology is perhaps due at this stage to those readers who would
have preferred a detailed discussion of the evidence from final-state
measurements in favour of (or against) QCD as \emph{the} theory of
hadronic physics. I rather took the liberty of reinterpreting the
title as `Tests and perspectives of our \emph{understanding} of QCD
through final-state measurements.'  Such tests are vital if we are to
extend the domain of confidence of our predictions, as has been
discussed in the cases of diffraction and power corrections.

The tests of course should be well thought through:
some considerations that come out of the still to be fully understood
$b$-excess story are (a) the importance (as ever) of quoting results
at hadron level, not some ill-defined parton level; and (b) that if
carrying out a test at a given level of precision (\eg NLO), it is
necessary that all stages of the theoretical calculation (including
for example the determination of the fragmentation function), be
carried out at that same level of precision.

Another, general, consideration is the need for the Monte Carlo models
to be reliable and accurate, whether they be used to reconstruct data
or to estimate backgrounds. This is especially relevant in cases where
the actual measurements are limited to corners of phase space or where
large extrapolations are needed. In this context the recent advances
in the extension of Monte Carlo models to NLO accuracy is a
significant development, and in the medium term we should expect
progress from the current `proof-of-concept' implementations to a
widespread availability of NLO-merged event generators.

To conclude, it could well be that a few years from now, many of the
measurements and theoretical approaches discussed here will have made
it to textbooks as `standard' QCD. We look forward to future speakers
on this topic have an equally varied (but different) range of `until
recently controversial' tests of QCD to discuss!

%----------------------------------------------------------------------
\section*{Acknowledgements}

I wish to thank Matteo Cacciari, John Collins, Yuri Dokshitzer,
Stefano Frixione, Hannes Jung, and Frank Schilling for numerous
helpful suggestions, discussions and comments during the preparation
and writeup of this talk.

%----------------------------------------------------------------------

\end{document}